\documentclass{llncs}
\usepackage{amsfonts}
\usepackage{latexsym}
\usepackage{amssymb}
\usepackage[dvips]{graphicx}
\input{epsf}

\newcommand{\Sy}{{\cal S}}	
\newcommand{\Rf}{{\cal R}}	
\newcommand{\slts}{{\cal T}}	
\newcommand{\interface}{\mathit{Interface}}
\newcommand{\traces}{\mathit{Traces}}
\newcommand{\paths}{\mathit{Paths}}
\newcommand{\proj}{\mathit{Proj}}
\newcommand{\estdef}{~~\widehat{=}~~}
\newcommand{\guard}{\mathit{Guard}}
\newcommand{\action}{\mathit{Action}}
\newcommand{\Init}{\mathit{Init}}
\newcommand{\II}{{\cal I}}
\newcommand{\trc}{\rightsquigarrow}

\newcommand{\genesyst}{{\sf GeneSyst}}

\newcommand{\demoney}{{\sf Demoney}}   

\newcommand{\GetData}{\mathit{GetData}}
\newcommand{\InitializeTransaction}{\mathit{InitializeTransaction}}
\newcommand{\CompleteTransaction}{\mathit{CompleteTransaction}}
\newcommand{\Reset}{\mathit{Reset}}
\newcommand{\Error}{\mathit{Error}}
\newcommand{\TypeTransactions}{\mathit{TransactionType}}
\newcommand{\TypesOfStatus}{\mathit{StatusType}}
\newcommand{\TransEngage}{\mathit{EngagedTrans}}
\newcommand{\StatusWord}{\mathit{StatusWord}}
\newcommand{\CurTransaction}{\mathit{CurTransaction}}
\newcommand{\ChannelIsSecured}{\mathit{ChannelIsSecured}}

\newtheorem{defin}{Definition}
\newtheorem{lemme}{Property}
\newtheorem{theo}{Theorem}
\newtheorem{condi}{Condition}

\begin{document}


\newcommand{\B}{{\sf B}}		
\newcommand{\ab}{{\it AtelierB}}	


\newcommand{\incl}{{\sc includes}}	
\newcommand{\uses}{{\sc uses}}		
\newcommand{\sees}{{\sc sees}}		
\newcommand{\impo}{{\sc imports}}	
\newcommand{\prom}{{\sc promotes}}	
\newcommand{\exte}{{\sc extends}}	
\newcommand{\refines}{{\sc refines}}	


\newcommand{\mach}{{\sc machine}}	
\newcommand{\refi}{{\sc refinement}}	
\newcommand{\impl}{{\sc implementation}}
\newcommand{\bend}{{\sc end}}		
\newcommand{\ctrs}{{\sc constraints}}	
\newcommand{\defs}{{\sc definitions}}	
\newcommand{\sets}{{\sc sets}}		
\newcommand{\cons}{{\sc constants}}	
\newcommand{\prop}{{\sc properties}}	
\newcommand{\var}{{\sc variables}}	
\newcommand{\inv}{{\sc invariant}}	
\newcommand{\assert}{{\sc assertions}}	
\newcommand{\init}{{\sc initialisation}}
\newcommand{\oper}{{\sc operations}}	
\newcommand{\abscons}{{\sc abstract\_constants}}	
\newcommand{\concons}{{\sc concrete\_constants}}	
\newcommand{\absvar}{{\sc abstract\_variables}} 	
\newcommand{\convar}{{\sc concrete\_variables}} 	
\newcommand{\values}{{\sc values}}	


\newcommand{\bassert}{{\sc assert}}	
\newcommand{\bbegin}{{\sc begin}}	
\newcommand{\bpre}{{\sc pre}}		

\newcommand{\bif}{{\sc if}}		
\newcommand{\bthen}{{\sc then}}		
\newcommand{\belse}{{\sc else}}		
\newcommand{\belsif}{{\sc elsif}}	
\newcommand{\bcase}{{\sc case}}		
\newcommand{\bof}{{\sc of}}		
\newcommand{\beither}{{\sc either}}	
\newcommand{\bor}{{\sc or}}		
\newcommand{\bselect}{{\sc select}}	
\newcommand{\bwhen}{{\sc when}}		
\newcommand{\bchoice}{{\sc choice}}	

\newcommand{\bvar}{{\sc var}}		
\newcommand{\bany}{{\sc any}}		
\newcommand{\bwhere}{{\sc where}}	
\newcommand{\bin}{{\sc in}}		
\newcommand{\blet}{{\sc let}}		
\newcommand{\bbe}{{\sc be}}		
\newcommand{\bwhile}{{\sc while}}	
\newcommand{\bdo}{{\sc do}}		
\newcommand{\bvariant}{{\sc variant}}	
\newcommand{\brec}{{\sc rec}}           


\newcommand{\bbool}{\mathsf{bool}}	
\newcommand{\bskip}{\mathsf{skip}}	



\newcommand{\lneg}{\neg\,}		
\newcommand{\logand}{~\wedge~}		
\newcommand{\logor}{~\vee~}		
\newcommand{\logimpli}{\Rightarrow}	
\newcommand{\logequiv}{\Leftrightarrow}	
\newcommand{\T}{\mathsf{btrue}}		
\newcommand{\F}{\mathsf{bfalse}}	


\newcommand{\bguard}{\Longrightarrow}	
\newcommand{\bch}{~[\!]~}		
\newcommand{\ba}{@}		        
\newcommand{\refs}{\sqsubseteq}		
\newcommand{\bpt}{\cdot}		
\newcommand{\notfree}{\backslash}	
\newcommand{\bres}{\longleftarrow}	
\newcommand{\pv}{~\mathbf{;}~}	        


\newcommand{\btrm}{\mathsf{trm}}	
\newcommand{\babt}{\mathsf{abt}}	
\newcommand{\bmir}{\mathsf{mir}}	
\newcommand{\bfis}{\mathsf{fis}}	
\newcommand{\bgrd}{\mathsf{grd}}	
\newcommand{\bprd}{\mathsf{prd}}	

\newcommand{\rpre}{\mathsf{pre}}	
\newcommand{\rrel}{\mathsf{rel}}	
\newcommand{\rfnc}{\mathsf{fnc}}	
\newcommand{\rdom}{\mathsf{dom}}	
\newcommand{\rran}{\mathsf{ran}}	
\newcommand{\rstr}{\mathsf{str}}	
\newcommand{\rset}{\mathsf{set}}	
\newcommand{\rid}{\mathsf{id}}	        
\newcommand{\rprj}{\mathsf{prj}}	


\newcommand{\domr}{\ \mbox{{\large $\lhd$}}\ }      
\newcommand{\adomr}{\ \mbox{{\large $\lhd\!\!\!\!\!-$}}\ }  
\newcommand{\ranr}{\ \mbox{{\large $\rhd$}}\ }      
\newcommand{\aranr}{\ \mbox{{\large $\rhd\!\!\!\!\!-$}}\ }  
\newcommand{\bover}{\mbox{\ {\large $\lhd\!\!\!-$}}\ } 

\newcommand{\bpow}{{\mathbb P}} 	
\newcommand{\bpownv}{{\mathbb P}_1} 	
\newcommand{\bfin}{{\mathbb F}} 	
\newcommand{\bfinnv}{{\mathbb F}_1} 	

\newcommand{\brel}{\leftrightarrow}	
\newcommand{\funtot}{\rightarrow}	
\newcommand{\funpar}{\rightarrow\!\!\!\!\!\shortmid~~}	
\newcommand{\injtot}{\rightarrowtail}   
\newcommand{\injpar}{\rightarrowtail\!\!\!\!\!\shortmid~~} 
\newcommand{\surtot}{\twoheadrightarrow} 
\newcommand{\surpar}{\twoheadrightarrow\!\!\!\!\!\shortmid~~} 
\newcommand{\bijtot}{\rightarrowtail\!\!\!\!\!\twoheadrightarrow}
\newcommand{\bijpar}{\rightarrowtail\!\!\!\!\!\twoheadrightarrow\!\!\!\!\!\shortmid~~}                                 

\newcommand{\bunion}{\mathsf{union}}	
\newcommand{\binter}{\mathsf{inter}}	


\newcommand{\N}{{\mathbb N}}		
\newcommand{\Z}{{\mathbb Z}}		
\newcommand{\nat}{\mathsf{NAT}}	        
\newcommand{\natpos}{\mathsf{NAT}_1}    
\newcommand{\intrel}{\mathsf{INT}}	
\newcommand{\minint}{\mathsf{minint}}	
\newcommand{\maxint}{\mathsf{maxint}}	

\newcommand{\mod}{\mathsf{mod}}	
\newcommand{\bsucc}{\mathsf{succ}}	
\newcommand{\bpred}{\mathsf{pred}}	
\newcommand{\card}{\mathsf{card}}	
\newcommand{\bmin}{\mathsf{min}}	
\newcommand{\bmax}{\mathsf{max}}	

\newcommand{\caract}{\mathsf{CHAR}}	
\newcommand{\chaine}{\mathsf{STRING}}	

\newcommand{\bool}{\mathsf{BOOL}}	
\newcommand{\true}{\mathsf{TRUE}}	
\newcommand{\false}{\mathsf{FALSE}}	

\newcommand{\seq}{\mathsf{seq}}	        
\newcommand{\iseq}{\mathsf{iseq}}	
\newcommand{\perm}{\mathsf{perm}}	

\newcommand{\sv}{[\,]}	                
\newcommand{\insg}{~-\!\!\!\!>}		
\newcommand{\insd}{<\!\!\!\!-~}		
\newcommand{\sapp}{~^{\frown}~}		
\newcommand{\conc}{\mathsf{conc}}	
\newcommand{\shaut}{\uparrow}		
\newcommand{\sbas}{\downarrow}		
\newcommand{\size}{\mathsf{size}}	
\newcommand{\first}{\mathsf{first}}	
\newcommand{\last}{\mathsf{last}}	
\newcommand{\front}{\mathsf{front}}	
\newcommand{\tail}{\mathsf{tail}}	
\newcommand{\rev}{\mathsf{rev}}	        


\newcommand{\bigor}{\mathsf{OR}}	
\newcommand{\dirp}[1]{\otimes_{#1}}	



\def\machinebox#1{\centerline{\hbox{\fbox{\parbox[t]{4cm}{
	\vspace*{-2ex}
	\begin{tabbing}
	X\=XX\=XX\=XX\=XX\=XX\=XX\=XX\=XX\=XX\=XX\=XX\=XX\= \kill
	#1
	\end{tabbing}
	\vspace*{-4ex}
	}}}}
}


\def\machinesbox#1{\hbox{\fbox{\parbox[t]{4cm}{
        \vspace*{-2ex}
        \begin{tabbing}
        X\=XX\=XX\=XX\=XX\=XX\=XX\=XX\=XX\=XX\=XX\=XX\=XX\= \kill
        #1
        \end{tabbing}
        \vspace*{-4ex}
        }}}
}



\title{\genesyst: a Tool to Reason about \\Behavioral Aspects of \B\ Event 
Specifications.\\Application to Security Properties\thanks{This work 
was done in the GECCOO project of program ``ACI~: 
S\'ecurit\'e Informatique'' supported by the French Ministry of Research 
and New Technologies. It is also suported by CNRS and ST-Microelectronics 
by the way of a doctoral grant.}}

\titlerunning{\genesyst: a Tool to Reason about Event-\B\ 
Specifications}

\author{Didier Bert \and Marie-Laure Potet \and Nicolas Stouls}
\institute{
   Laboratoire Logiciels Syst\`emes R\'eseaux - LSR-IMAG - Grenoble, France
\\ 
   \email{\{Didier.Bert, Marie-Laure.Potet, Nicolas.Stouls\}@imag.fr}}

\maketitle

\begin{abstract}
In this paper, we present a method and a tool to build symbolic labelled
transition systems from \B\ specifications. The tool, called \genesyst,
can take into account refinement levels and can visualize the 
decomposition of abstract states in concrete hierarchical states.
The resulting symbolic transition system represents all the behaviors of the
initial \B\ event system. So, it can be used to reason about them.
We illustrate the use of \genesyst\ to check security
properties on a model of electronic purse.
\end{abstract}

\section{Introduction}
\label{intro}

Formal methods, such as the \B\ method \cite{BBook}, 
ensure that the development 
of an application is reliable and that properties expressed in the
model are satisfied by the final program. However, they do not 
guarantee that this program fulfills the informal requirements, 
nor the needs of the customer. So, it is useful to propose several 
views about the specifications, in order to be sure that the initial 
model is suitable for the customer and that the development 
can continue on this basis. One of these important insights is the
representation of the behavior of programs by means of diagrams
(statecharts). Moreover, some particular views, if they are themselves
formal, can provide new means to prove properties that cannot easily be
checked in the first model.

In this paper, we present a method and a tool to extract a labelled
transition system from a model written in event-\B. The transition
system gives a graphical view and represents symbolically all the 
behaviors of the \B\ model. The method is able to take into account
refinement levels and to show the correspondence between
abstract and concrete systems, by means of hierarchical states.

We present also an application of this tool, namely, the
verification of security properties. 
The security properties assert the occurrence or 
the absence of some particular events in some situation. 
They are a case of {\em atomicity property} of transactions.
This is illustrated by an example
of specification of an electronic purse, 
called \demoney \cite{DEMONEY-Spec,DEMONEY-AnnexeDetaillees},
developed in the SecSafe project \cite{SecSafe}. 
This case study, written in Java Card \cite{APIJavaCard}, is an applet 
that has all the facilities required by 
a real electronic purse. Indeed, the purse can be debited 
from a terminal in a shop, credited by cash or from a 
bank account with a terminal in a bank or managed from special 
terminal in bank restricted area.
Transactions are encrypted if needed and different 
levels of security are used depending on the actions. 
\demoney\ also supports to communicate with another 
applet on the card, for example, to manage award points on a loyalty plan. 
The specification of \demoney\ is public in 
version 0.8 \cite{DEMONEY-Spec}, but the source code is copyrighted 
by Trusted Logic S.A.\footnote{http://www.trusted-logic.fr/}.

In Section~\ref{eventB}, we recall the main features of
event-\B\ systems and refinements. We introduce a notion of behavioral
semantics by the way of sequences of events. In Section~\ref{LTSconstruction},
we define symbolic labelled transition systems (SLTS) and the links between
SLTS and event-\B\ systems are stated. 
In Section~\ref{genesyst-tool}, we present
the \genesyst\ tool and an example of generation of SLTS dealing with the
error cases in the \demoney\ case study. Section~\ref{verif-security} 
presents security properties required in the application 
and shows how the \genesyst\ diagrams can be used to
check these properties. Then, we review related works, and we conclude 
the paper with some
research perspectives in Section~\ref{conclusion}.

\section{Event-\B}
\label{eventB}

\subsection{General presentation}
\label{eventB-presentation}

Event-\B\ was introduced by J.-R. Abrial \cite{abrial96,JRAbrialB98}.
It is a formal development method as well as a specification language.
In event-\B, components are composed of constant declarations 
(\sets, \cons, \prop),
state specification (\var, \inv), initialisation and set of {\em events}.
The events are defined
by $e \estdef \mathit{eBody}$ where $e$ is the name of the event
and $\mathit{eBody}$ is a {\em guarded} generalized substitution \cite{BBook}.
The events do not take parameters and do not return result values.
They do not get preconditions and do terminate. Their effect is only 
to modify the internal state. 
If $\Sy$ is a component, then we 
denote by $\interface(\Sy)$ the set of its events.

A well-typed and well-defined component 
is consistent if initialization $\Init$ establishes the
invariant of the component and if each 
event preserves the invariant.
So, using the notation $[S]R$ as the weakest precondition of $R$ for 
substitution $S$, the consistency of a component is expressed by the 
proof obligations: $[\Init]I$ and $I \logimpli [\mathit{eBody}]I$ for each
event.

In the paper, we use the notions of before-after predicate 
of substitution $T$ for variables $x$ ($\bprd_x(T)$) and the 
feasability predicate of a substitution as defined in
the B-Book: $\bfis(T)\logequiv \neg[T]\mathit{false}$ \cite{BBook}. 
Finally, the notation $\langle T \rangle R$
means $\neg [T] \neg R$, that is to say, there exists a computation of 
$T$ which terminates in a state verifying $R$.

\subsection{Events and traces}
\label{events-traces}

The events have the form ``$e \estdef G \bguard T$'' where 
$G$ is a predicate, $T$ is a generalized substitution such that 
$I \land G \logimpli \bfis(T)$.
Predicate $G$ is called the {\em guard} of $e$ and $T$ is its {\em action}.
They are respectively denoted by $\guard(e)$ and $\action(e)$. 
If the syntactic definition of an event $e \estdef S$ does not fulfill
this form, it can be built by computing $e \estdef \bfis(S)\bguard S$.
Following the so-called {\em event-based} approach \cite{csp}, 
the semantics of event-\B\ systems can be chosen to be the set of all 
the valid sequences of event executions. 

\begin{defin}[Traces of Event-\B\ systems]
\label{def-traces}
A finite sequence of event occurrences $e_0.e_1.e_2\ldots e_n$
is a trace of system $\Sy$ if and only if $e_0$ is the initialisation
of $\Sy$, $\{e_1,e_2,\ldots,e_n\} \subseteq \interface(\Sy)$ and
$\bfis(e_0\pv e_1\pv e_2\pv\ldots \pv e_n) \logequiv true$.
\end{defin}
The set of all the finite traces of a system $\Sy$ is called $\traces(\Sy)$.
For the initialisation, one can notice that $\bprd_x(\Init)$ does not 
depend on the initial values of the variables and that 
$\guard(\Init)\logequiv true$. The following property characterizes traces
by the existence of intermediary states $x_i$ in which the guard of $e_i$
holds and where the pair $(x_i,x_{i+1})$ is in the before-after predicate 
of event $e_i$:

\begin{lemme}[Trace characterization]
\label{charac-tr}
Let $x$ be the variable space of system $\Sy$, then: ~~~~~$e_0.e_1.\ldots e_n \in \traces(\Sy) ~~\logequiv$\\
$\begin{array}{l}
~~~\exists x_0,\ldots,x_{n+1}\bpt\bigwedge_{i=0}^n ([x:=x_i]\guard(e_i)\land[x,x':=x_i,x_{i+1}]\bprd_x (\action(e_i)))
\end{array}$
\end{lemme}

\subsection{Event-\B\ refinement}
\label{eventB-ref}

In the event-\B\ method, a refinement is a component called \refi. 
The variables
can be refined (i.e. made more concrete) and a {\em gluing invariant}
describes the relationship between the variables of the refinement and those
of the abstraction. The events of refinement $\Rf$ must at least contain
those of the abstraction $\Sy$ (i.e. $\interface(\Sy) \subseteq 
\interface(\Rf)$). The other events are called {\em new} events.

We recall here the proof obligations of system refinements. Let $I$ be
the invariant of the abstraction $\Sy$ and $J$ be the invariant of
refinement $\Rf$, then the gluing invariant is the conjunction $I \land J$.
%
%
The refinement is performed elementwise, 
that is to say, the abstract initialisation is refined by the concrete 
initialisation and each abstract event is refined by its concrete 
counterpart. Proof obligations that establish the consistency of 
refinements are~:

\medskip
\begin{tabular}{ll}
For initialisation $\Init$~:  & $[\Init^R]\langle \Init^S \rangle J$\\
For events $e$ of $\interface(\Sy)$~:~~~ & $I \land J \logimpli [e^R]\langle e^S \rangle J$\\
For the new events $ne^R$~: & $I \land J \logimpli [ne^R]\langle \bskip \rangle J$\\
\end{tabular}

\medskip
\noindent
New events cannot indefinitely take the control, i.e. the refined 
system cannot diverge more often that the abstract one. So, a {\em variant}
$V$ is declared in the refined system, as an expression on a well-founded
set (usually the natural numbers), and the new events must satisfy ($v$ is a
fresh variable)~:

\medskip
\begin{tabular}{ll}
$V$ is a natural expression~:  & $I \land J \logimpli V \in \N$\\
New events $ne^R$ decrease the variant~:~~~ & $I \land J \logimpli [v:=V][ne^R](V < v)$
\end{tabular}

\medskip
\noindent
Finally, a proof obligation of {\em liveness preservation} is usually required.
If $\Sy$ contains $m$ events and $\Rf$ contains $p$ new events, then:

\medskip
\begin{tabular}{l}
$I \land J \logimpli (\bigvee_{i=1}^{m} \guard (e^S_i) \logimpli (\bigvee_{i=1}^{m} \guard(e^R_i) \logor \bigvee_{i=1}^{p} \guard(ne^R_i)))$
\end{tabular}

\medskip
\noindent
Traces associated to refinements are defined as for the systems.


\section{Symbolic labelled transition systems associated to \B\ systems}
\label{LTSconstruction}

\subsection{Symbolic transition systems}
\label{symbolic-TS}

We define {\em symbolic} labelled transition systems: 

\begin{defin}[Symbolic labelled transition system]
\label{def-lts}
A symbolic labelled transition system (SLTS) is a 4-uple $(N, \Init, U, W)$ where\\
\begin{tabular}{l}
- $N$ is a set of states, and $\Init$ is the initial state $(\Init \in N)$\\
- $U$ is a set of labels of the form $(D,A,e)$, where $D$ and $A$
are predicates and\\
~~~~ $e$ is an event name\\
- $W$ is a transition relation $W \subseteq \bpow(N \times U \times N)$.
\end{tabular}
\end{defin}

A transition $(E,(D,A,e),F)$ means that, in state $E$, the event $e$ 
is enabled if $D$ holds and, starting from
state $E$, if event $e$ is enabled, then it reaches state $F$ if $A$ holds. 
Predicate $D$ is called the {\em enabledness}
predicate and $A$ is called the {\em reachability} predicate.

States $N$ are interpreted as subsets of variable spaces on 
variables $x$. So, the interpretation of $N$ is given by a function $\II$
such that $\II(E)$ is a predicate on free variables $x$ which characterizes
the subset represented by $E$.
In the next definition, we determine the actual conditions to cross
a transition from a particular state value $x_1$ of $E_1$ to $x_2$ of $E_2$
by an event $e$ which is defined in an event-\B\ system $\Sy$.
For that, $e$ must be enabled in $x_1$,
$x_2$ must be reachable from $x_1$ by $e$, 
and ($x_1,x_2)$ must belong to the before-after predicate of $e$:

\begin{defin}[Transition crossing]
\label{def-tr-crossing}
Let $(E_1,(D,A,e),E_2)$ be a transition of a SLTS $\slts$ 
on a system $\Sy$, and given $x_1$ and $x_2$ 
some values of the state variables $x$ which satisfy the invariant of $\Sy$, 
then a crossing from $x_1$ to $x_2$ by this transition 
is legal if and only if~:
\begin{tabular}[t]{l}
1.~ $[x:=x_1](\II(E_1) \land D \land A)$\\
2.~ $[x,x' := x_1,x_2]\,\bprd_x(\action(e))$\\
3.~ $[x:=x_2]\II(E_2)$\\
\end{tabular}\\
Such a legal transition crossing is denoted by~:
\[ (E_1,x_1) \trc^{(D,A,e)}\!\trc~(E_2,x_2)  \]
\end{defin}

Now, we introduce the notion of {\em path} in a symbolic labelled transition 
system.
A path is a sequence of event occurrences, starting from the initial 
state, which goes over a transition system through legal transition 
crossings.

\begin{defin}[Paths]
\label{def-path}
Given a symbolic labelled transition system $\slts$ on a system $\Sy$,
a sequence of event occurrences $e_0.\ldots.e_{n+1}$ is a path in 
$\slts$
if there exists a list of states $E_0,\ldots,E_{n+1}$ of $N$, with
$E_0=Init_{\slts}$, and a
list of transitions $(D_i,A_i,e_i), i\in 0..n$, such that~:~~\\
\begin{tabular}{l}
~~~~~~$\exists x_0,\ldots,x_{n+1}\bpt(\bigwedge_{i=0}^n ((E_i,x_i)\trc^{(D_i,A_i,e_i)}\!\trc~(E_{i+1},x_{i+1})))$
\end{tabular}
\end{defin}
The set of all the finite paths of $\slts$ is called 
$\paths(\slts)$.

\subsection{Construction of states and transitions}
\label{statesC}

The aim of this section is to show how to
compute a SLTS, from an event-\B\ system $\Sy$ and given a set of states $N$.
First, to build the states $N$, consider a list of predicates 
$\{P_1,\ldots,P_n\}$ on the variable space. 
We require that this set is {\em complete} 
with respect to the invariant, i.e. all the states specified by the 
invariant are included in the states determined by the $P_i$ 
predicates, i.e.
\[ I ~\logimpli~ \bigvee_{i=1}^n P_i \]
Then, the states of the SLTS are $N = \{\Init_\Sy, E_1,\ldots,E_n\}$ 
with the interpretation defined by:

\medskip
\begin{tabular}{ll}
$\II(\Init_\Sy) = true$ ~~~~~~~~~~~~~~~~& $\II(E_i) = P_i \logand I,~~~ i \in 1..n$\\
\end{tabular}

\medskip
\noindent
We denote by $N1$ the set $N - \{\Init_\Sy\}$. From the completeness property 
above and the definition of $N$, we get:
$I ~\logequiv~ \bigvee_{i=1}^n \II(E_i)$.


Now, we express the conditions to ensure that
a symbolic labelled transition system $\slts$ represents the same set of 
behaviors as the associated system $\Sy$. For that, in a starting state $E$,
the enabledness condition must be equivalent to the guard of the event $e$, 
and if the target state is $F$, the reachability condition
must be equivalent to the possibility to reach $F$ through $e$, 
when the enabledness predicate holds, so the condition:

\begin{condi}[Valid transitions]
\label{def-tr-validity}
Let $\Sy$ be a system, $E$ and $F$ two states in $N$ as defined above, 
and $e$ an event, then the transition (E,(D,A,e),F) is valid 
if and only if  predicates $D$ and $A$ satisfy~:\\
\begin{tabular}{lll}
~~~~& $a)$~~~~~& $\II(E) ~\logimpli~ (D \logequiv \guard(e))$\\
& $b)$ & $\II(E) \land \guard(e)~\logimpli~ (A \logequiv \langle\action(e)\rangle \II(F))$\\
\end{tabular}
\end{condi}
Notice that, by applying the definition of the conjugate weakest precondition,
condition~$b)$ is equivalent to~:
\[ \II(E) \land \guard(e)~\logimpli~ (A \logequiv \exists x'\bpt(\bprd_x(\action(e)) \land [x:=x']\II(F))) \]
%
A SLTS with all the transitions valid with respect to a system $\Sy$ is
called a valid symbolic labelled transition system.

\begin{theo}[Traces and paths equality]
\label{tp-equality}
Let $\Sy$ be an event-\B\ system with invariant $I$ and events $Ev$
and let $\slts$ be a valid symbolic labelled transition system  
built from $\Sy$, then:
\[ \traces(\Sy) = \paths(\slts) \]
\end{theo}

\noindent
{\bf Proof:} We prove that, for all $t$, $t \in \paths(\slts) \logequiv t \in \traces(\Sy)$.\\
The path $t \estdef e_0.e_1.\ldots.e_n$ is a path for the state sequence $E_0,E_1,\ldots,E_{n+1}$ iff (Definition~\ref{def-path}):
\begin{tabular}{c}
~~~~$\exists x_0,\ldots,x_{n+1}\bpt\bigwedge_{i=0}^n ((E_i,x_i)\trc^{(D_i,A_i,e_i)}\!\trc~(E_{i+1},x_{i+1}))$\\
\end{tabular}\\
By using Definition~\ref{def-tr-crossing}, we get:\\
\begin{tabular}{c}
~~~~$\exists x_0,\ldots,x_{n+1}\bpt\bigwedge_{i=0}^n ([x:=x_i](\II(E_i) \land D_i \land A_i)$\\ 
~~~~~~~~$\logand [x,x':=x_i,x_{i+1}]\bprd_x(\action(e_i)) \logand [x:=x_{i+1}]\II(E_{i+1}))$\\
\end{tabular}\\
By Condition~\ref{def-tr-validity}, one can replace $D_i$
by $\guard(e_i)$ and $A_i$ by $\exists x'\bpt(\bprd_x(\action(e_i)) \land [x:=x']\II(E_{i+1}))$. The formula above is simplified and becomes:\\
\begin{tabular}{lc}
(1)~~~~~& $\exists x_0,\ldots,x_{n+1}\bpt\bigwedge_{i=0}^n ([x:=x_i](\II(E_i) \land \guard(e_i))$\\ 
& $\logand [x,x':=x_i,x_{i+1}]\bprd_x(\action(e_i)) \logand [x:=x_{i+1}]\II(E_{i+1}))$\\
\end{tabular}\\
We must prove that this formula is equivalent to the characterization 
of the traces (Property~\ref{charac-tr}):\\
\begin{tabular}{lc}
(2)~~~~~& $\exists x_0,\ldots,x_{n+1}\bpt\bigwedge_{i=0}^n ([x:=x_i]\guard(e_i)$\\
& $\logand [x,x':=x_i,x_{i+1}]\bprd_x (\action(e_i)) \logand [x:=x_{i+1}]I)$
\end{tabular}\\
Implication $(1) \logimpli (2)$ is verified because states $E_i$ are such 
that $\II(E_i) \logimpli I$ (Section~\ref{statesC}).
To prove $(2) \logimpli (1)$, we must exhibit a list of states 
$E_0, E_1,\ldots,E_{n+1}$ such that these states satisfy (1). This
follows from the fact that $\II(E_0) = true$ and from 
$I \logimpli \bigvee_{i=1}^n \II(E_i)$, which ensures that 
one of the states $\II(E_i)$ necessarily holds when $I$ hold.\hfill$\Box$

\subsection{Labelled transition systems for the refinements}
\label{LTSrefinement}

We propose now the construction of a symbolic labelled transition system 
for the refinements. Our aim is to highlight the links between 
abstract and concrete transition systems, while preserving the overall 
structure of the abstract system. 
One aspect of the refinement is the change of the variable representation  
and redefinition of the events of the abstraction, according to the new
representation.
The point is taken into account by the notion of 
{\em state projection}.

In the following, $\Sy$ is a specification, $\Rf$ is its
refinement with gluing invariant $L$, and $\slts^S$ is a symbolic
labelled transition system for $\Sy$. States $E^S$ and $F^S$ are states 
in $\slts^S$.
We assume that the variable set $x^S$ of $\Sy$ is
disjoint to the variable set $x^R$ of the refinement. If some variables 
of the specification are
kept in the refinement, they can be renamed and an equality between
both variables is added to the invariant.

\begin{defin}[State projection]
\label{state-proj}
Let $\Sy$ be a system with variables $x^S$ and $\Rf$ be the 
refinement of $\Sy$ according to $L$.
A state $E^R$ of $\slts^\Rf$, $E^R \not= Init_\Rf$ is the projection
of $E^S$ of $\slts^\Sy$, denoted by $E^R = \proj_L(E^S)$, iff:
\[ \II(E^R) \logequiv \exists x^S\bpt(L \logand \II(E^S)) \]
\end{defin}

We propose to build a SLTS, called $\proj_L(\slts^S)$,
in which states are automatically deduced from abstract states and 
gluing invariant.
The SLTS projection $\proj_L(\slts^S)$ of the refinement $\Rf$ of system $\Sy$
with gluing invariant $L$ is such that:
the initial state is any $q_0$ with $\II(q_0) = true$; 
the other states of the projection are the projections of abstract states,
i.e. $N1^R = \{\proj_L(q)~|~q \in N1^S\}$.
The transitions are $(E^R,(D',A',e^R),F^R)$ where
$e^R \in \Rf$ and $D'$, $A'$ are such that Condition \ref{def-tr-validity} 
is satisfied.
%
A transition $(E^R,(D',A',e^R),F^R)$ is said a {\em projection of transition}
$(E^S,(D,A,e^S),F^S)$ iff $E^R=\proj_L(E^S)$, $F^R=\proj_L(F^S)$ and event 
$e^R$ is the refinement of $e^S$.
By construction, 
$\paths(\proj_L(\slts^S)) = \traces(\Rf)$. 
This equality can be proved in the same way as 
in Theorem~\ref{tp-equality}.

\begin{lemme}[Transition projection]
\label{transition-proj}
With the definitions above,\\
let $(E^R,(D',A',e^R),F^R)$ be the projection of 
transition $(E^S,(D,A,e^S),F^S)$, then we have:
\[   \II(E^S) \logand L \logand D' \logimpli D  \]
\end{lemme}

This property says that any transition enabled from a state
$\proj_L(E^S)$ in a refinement $\Rf$
actually must be enabled in specification $\Sy$ (if the refinement is proved
correct). Property~\ref{transition-proj} can make the computation of the
transitions simpler. Indeed, 
if $e \in \interface(\Sy)$, then, for all the transitions $(E^S,e,F^S)$
of the abstraction, it is only necessary to examine the transitions
$(\proj_L(E^S),e,E')$ with $E' \in N1^R$. No other transition can be
labelled by $e$ from this state.


Another key aspect of refinement is the refinement of behaviors. 
New events may be introduced that make the
actions more detailed. These new events are not observable
at the abstract level, as the stuttering in TLA \cite{lamport94}. 
Very often, new variables are introduced. Thus, it is useful 
to visualize the states referring to these variable changes. 
In order to preserve the structure of the abstract system, we choose 
to refine each abstract state in an independent way. So, the
transitions, relative to events which belong to $\interface(\Sy)$,
are preserved by the introduction of hierarchical states.

\begin{defin}[Hierarchical states]
\label{H-States}
A set of sub-states 
$\{E^R_1,\ldots,E^R_m\}$ can be associated to a super-state
$\proj_L(E^S)$ of $\Rf$ if and only if
\[  \bigvee_{i=1}^m \II(E^R_i) \logequiv \II(\proj_L(E^S)) \]
\end{defin}

In a refined system, the user must decide what projections
of abstract states are decomposed and s/he must provide the predicates of 
the decomposition. If the abstract states are disjoint, then the transitions 
associated to the new events appear only between the sub-states of 
a hierarchical state. An example of refinement with decomposition of
states is given in Section~\ref{ts-demoney-2}.

\section{The \genesyst\ tool}
\label{genesyst-tool}

\subsection{Presentation}
\label{genesyst-presentation}

The \genesyst\ tool is intended to generate a symbolic 
labelled transition system $\slts$ from an event-\B\ system $\Sy$ 
and a set of states $N$.
Such a generated SLTS will be denoted by $\slts(\Sy,N)$.
The input of the tool is a \B\ component,
where the \assert\ clause contains the formula $P_1 \lor \ldots \lor P_n$,
which characterizes the list of predicates $\{P_1,\ldots,P_n\}$.
By this way, the condition of completeness (section~\ref{statesC})
is generated as proof obligation.

%
%

We give a sketch of the algorithm which computes the transitions of 
$\slts(\Sy,N)$: it uses three main variables:
the set of visited states, $visited$, the set of processed states,
$processed$, and the set of computed transitions $tr$.
First, the initial state is put in the $visited$ set. Then
each state $E$ in the $visited$ set is processed: this consists in computing
the transitions $(E,(D,A,e),F)$ with all events $e$ to all
non-initial states $F$ of the system. Predicates $D$ and $A$ are 
determined following the algorithm defined in the following section.
If $D$ or $A$ are not $\mathit{false}$ then the transition
$(E,(D,A,e),F)$ is added to $tr$, 
and if $F$ has not been processed, it is put in the $visited$ set. 
After the processing of state $E$, $E$ is removed from $visited$
and put in set $processed$. When $visited$ is empty, then $tr$
contains all the computed transitions of $\slts(\Sy,N)$ and
$processed$ contains the set of reachable states. The algorithm
terminates, because the set of states to be visited is finite (bounded by
the cardinal of $N$). This algorithm guarantees that the resulting SLTS 
is a valid transition system for $\Sy$, with given states $N$.

\subsection{Proof obligations}
\label{genesyst-prf-obligations}

A subprocedure of the algorithm is to determine effectively 
the enabledness predicate and 
the reachability predicate, given a triple $(E,e,F)$.
For sake of usability of the resulting transition system, it is 
interesting to examine three cases: predicates are $true$, $\mathit{false}$
or other. This information can be obtained by proof obligations.
In Fig. \ref{po-D-A}, we give the conditions for 
the calculus of these predicates. Obviously, if $D$ and/or $A$
is $\mathit{false}$, then the transition is not possible.

\begin{figure}[htb]
\begin{center}
\begin{tabular}{|l|l|c|}
\hline
& \multicolumn{1}{|c|}{Proof obligations} & \multicolumn{1}{c|}{$D$ for $(E,e,F)$}\\
\hline
~(1)~~~~ & ~~$\forall x\bpt(\II(E) \logimpli \guard(e))$ & $true$\\
~(2) & ~~$\forall x\bpt(\II(E) \logimpli \neg \guard(e))$ & $\mathit{false}$\\
~(3) & ~~$\exists x\bpt(\II(E) \land \guard(e))$ & $\guard(e)$\\
\hline
& \multicolumn{1}{|c|}{Proof obligations} & \multicolumn{1}{c|}{$A$ for $(E,e,F)$}\\
\hline
~(4) & ~~$\forall x\bpt(\II(E) \land \guard(e) \logimpli \langle \action(e) \rangle \II(F))$ & $true$\\
~(5) & ~~$\forall x\bpt(\II(E) \land \guard(e) \logimpli [\action(e)] \neg\, \II(F))$~~~~~ & $\mathit{false}$\\
~(6) & ~~$\exists x\bpt(\II(E) \land \guard(e) \land \langle \action(e) \rangle \II(F))$ & ~~~$\langle \action(e) \rangle \II(F)$~~~\\
\hline
\end{tabular}
\vskip -6pt
\caption{Proof obligations for enabledness and reachability}
     \label{po-D-A}
\end{center}
\end{figure}

\vspace*{-3mm}
In practice, the \genesyst\ tool computes the proof obligations (POs) above
and interacts with AtelierB to discharge the POs. For each triple
$(E,e,F)$:
\begin{enumerate}
\item if proof obligation (1) is automatically discharged 
then $D$ is $true$.
\item if proof obligation (2) is automatically discharged 
then $D$ is $\mathit{false}$ and
transition $(E,e,F)$ does not occur in the resulting $\slts(\Sy,N)$.
\item otherwise, $D$ is $\guard(e)$ by default.
\end{enumerate}

\noindent
Then, after cases 1. and 3., \genesyst\ computes the proof obligations for 
determining the reachability predicate $A$.

\begin{enumerate}
\item[4.] if proof obligation (4) is automatically discharged 
then $A$ is $true$.
\item[5.] if proof obligation (5) is automatically discharged 
then $A$ is $\mathit{false}$ 
and transition $(E,e,F)$ does not occur in the resulting $\slts(\Sy,N)$.
\item[6.] otherwise, the transition is kept with 
$\langle \action(e) \rangle\, \II(F)$ as $A$, by default.
\end{enumerate}

We can notice that Condition~\ref{def-tr-validity} about the validity 
of the transitions is well satisfied by construction. 
The {\em by default cases} in 3. and 6. 
correspond to several possibilities. 
Either there exist values in state $E$ for which the transition is crossable
(guard of $e$ is true and state $F$ is reachable), or there are not
(the guard is false or state $F$ is not reachable).
However, in both possibilities, these transitions are included in the 
resulting transition system. 
To manage this feature, we define the notion of {\em minimal} 
symbolic labelled transition system.

\begin{defin}[Minimal SLTS]
\label{DefMinimality}
A minimal SLTS is a SLTS where all the transitions are valid, i.e. satisfy a) and b) of Condition 
\ref{def-tr-validity}, and also satisfy:\\
\begin{tabular}{lll}
~~~~& $c)$~~~~~& $D \not\logequiv \mathit{false}$ and $A \not\logequiv \mathit{false}$
\end{tabular}
\end{defin}

A SLTS built by \genesyst\ is minimal if all the proof obligations of
$D$ and $A$ have been effectively discharged in step 1. or 2. and 
step 4. or 5. in the algorithm above. To minimize the number of by-default
transitionss, we have designed two variants of the algorithm.
The first optional alternative of the algorithm is to change cases 3. and 6. 
into:

\begin{enumerate}
\item[3'.] if proof obligation (3) is automatically discharged, 
then $D$ is $\guard(e)$ by proof, otherwise, D is $\guard(e)$ by default.
\item[6'.] if proof obligation (6) is automatically discharged,
then $A$ is $\langle \action(e) \rangle\, \II(F)$ by proof, otherwise, 
the transition is kept with 
$\langle \action(e) \rangle\, \II(F)$ as $A$ by default.
\end{enumerate}

Another option of the tool allows the user to get the POs which have not been
automatically discharged. Then, s/he can do an interactive proof to
complete the work and return the information that the PO is discharged 
or not. However, the interactive mode is not very practicable when there
are a great number of proof obligations that are not automatically discharged.
It becomes useful to check actually the absence of some critical 
transitions (cases 2. and 4.).


\subsection{Transition systems associated to the \demoney\ case study}
\label{ts-demoney-1}

\begin{figure}[htb]
     \centerline{\epsfxsize=14cm \epsfbox{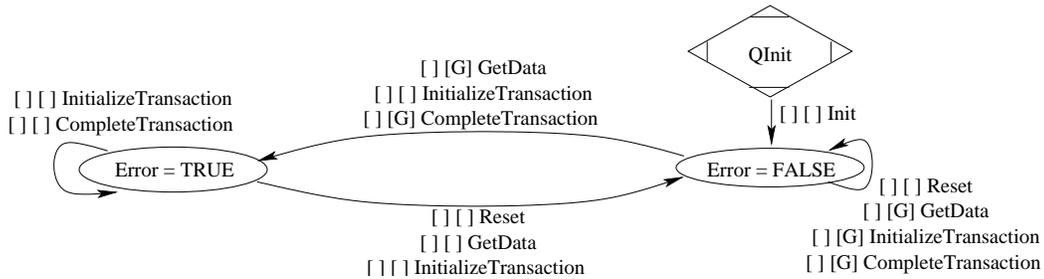}}
     \vskip -8pt
     \caption{Transition system associated to the error detection in the \demoney\ specification}
     \label{ex-litle-slts}
   \end{figure}

   In Fig.~\ref{ex-litle-slts}, we give an example of transition system 
generated from a subset of the abstract specification of the 
\demoney\ case study. The \B\ machine is provided in appendix.
We just have represented four methods imposed by the \demoney\ 
specification \cite{DEMONEY-Spec}: 
{\it InitializeTransaction}, {\it CompleteTransaction}, {\it Reset} and 
{\it GetData}. 
The two methods {\it InitializeTransaction} and {\it CompleteTransaction} 
have to be executed in sequence. If they 
are called in the wrong order then an error must be returned.
Moreover, any other methods cannot be invoked between them, except 
the method {\it Reset} which models the extraction of 
the card from the terminal. If it is called during a transaction, all 
the internal variables must be restored at their initial values. 
Finally, method {\it GetData} has been defined to represent any other
method which plays a neutral r\^{o}le with respect to transactions.
   
Let us notice that our model has been expressed with
events. 
In the applet \demoney, methods have neither parameters nor result, 
because they communicate through a global variable,
named {\it APDU}, which allows the information transfert between 
the card and the terminal. 
An error can be returned by means of the same variable.
Finally, methods have no precondition,
because they are callable at any time.
So the transformation of methods in events is straightforward.

In the diagrams generated by \genesyst, transitions
are prefixed by the information about predicates $D$ and $A$.
A predicate denoted by ``[~]'' means $true$, while ``$[G]$''
means that the transition is computed by cases 3. or 6. (see
section~\ref{genesyst-prf-obligations}). 

\vskip -8pt

   Fig.~\ref{ex-litle-slts} points out cases in which errors can occur.
Transitions have no
enabledness condition, because all the guards are $true$ in the model.
Some reachability conditions do not reduce to $true$, as
for the event {\it GetData}, which is defined by: 
  
\begin{small}   
   \centerline{
     \hbox{
       \parbox[t]{4cm}{
	 \vspace*{-2ex}
	 \begin{tabbing}
           X\=XX\=XX\=XX\=XX\=XX\=XX\=XX\=XX\=XX\=XX\=XX\=XX\= \kill
           $\GetData =$ \>\>\>\>\bif\ $\TransEngage=\true$ \bthen\ \\
           \>\>\>\>\>$\Error:=\true ~||~ \TransEngage:=\false$\\
           \>\>\>\>\belse\ $\Error:=\false$ \bend;\\
           ~ 
	 \end{tabbing}
         \vspace*{-4ex}
       }
     }
   }
\end{small}   

\noindent  
From state {\small $\Error=\false$}, event $\GetData$
can reach the state {\small $\Error=\true$} with the condition 
{\small $\TransEngage=\true$}
and stays in {\small $\Error=\false$} otherwise.
Let us remark also that $\GetData$ is enabled in
state {\small $\Error=\true$} and always reaches state {\small $\Error=\false$}
because of the invariant 
{\small $\Error=\true \logimpli \TransEngage=\false$}.

\subsection{Transition system associated to a refinement of \demoney}
\label{ts-demoney-2}

   In our refinement of \demoney, the boolean variable $\Error$ is
changed into a value of a given set $\TypesOfStatus$,
which intends to describe error codes, as imposed by the
specification \cite{DEMONEY-Spec}.
In the same way, the boolean variable $\TransEngage$ is refined into
a value of a given set $\TypeTransactions$, which indicates
the exact type of the current transaction.
Finally, we have
introduced the channel with two levels of security ({\small $\false$} and
{\small $\true$}). All this
information is declared in the invariant below
(see also the refinement in appendix):

\medskip
\begin{small}
\machinebox{
  \>\inv\\
  \>\>$\StatusWord \in \TypesOfStatus \logand \CurTransaction \in \TypeTransactions \logand$ \\
  \>\>$\ChannelIsSecured \in \bool \logand$ \\
  \>\>$((\Error=\false) \logequiv (\StatusWord=ISO\_Ok)) \logand$ \\
  \>\>$((\TransEngage=\false) \logequiv (\CurTransaction = None)) \logand$ \\
  \>\>$((\CurTransaction \not= None) \logimpli (\ChannelIsSecured=\true)) \logand$ \\
  \>\>$((\StatusWord \not=ISO\_Ok) \logimpli (\CurTransaction = None))$\\
}
\end{small}

\begin{figure}[htb]
     \centerline{\epsfxsize=14cm \epsfbox{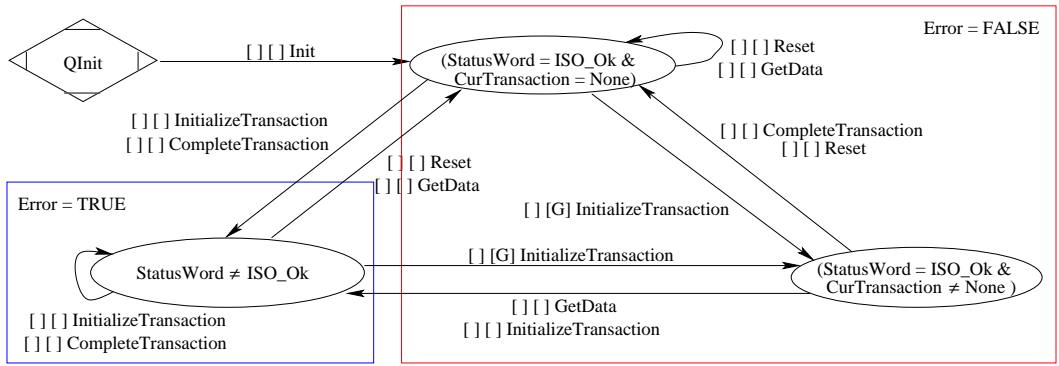}}
     \vskip -8pt
     \caption{Transition system associated to the refinement of the error detection}
     \label{ex-raf-slts}
   \end{figure}

\medskip
Fig.~\ref{ex-raf-slts} is built from this refinement.
State {\small $Error=\false$}, which corresponds to 
{\small $\StatusWord=ISO\_Ok$}, is
split into two states according to that a transaction
is engaged or not.

\vskip -8pt

As expressed in Definition~\ref{H-States}, the predicate given to \genesyst\ 
to describe the
states has to be a conjonction of equivalences between an abstract
state and a disjonction of refined states. This predicate is written
in the assertion clause. For example, the assertion below has 
been used to generate Fig.~\ref{ex-raf-slts}.

\begin{small}
\machinebox{
  \>$((\Error=\true) \logequiv$
  $((\StatusWord\not=ISO\_Ok \logand \CurTransaction=None)$ \\
  \>\>\>\>\>\>\>$\logor (\StatusWord\not=ISO\_Ok \logand \CurTransaction\not=None)))$\\
  \> $\logand $\\
  \>$((\Error=\false) \logequiv$
  $((\StatusWord=ISO\_Ok \logand \CurTransaction=None)$ \\
  \>\>\>\>\>\>\>$\logor (\StatusWord=ISO\_Ok \logand \CurTransaction\not=None)))~~$\\[-1mm]
}
\end{small}

\medskip
With the splitting of the state {\small $Error=\false$}, transition conditions 
are simplified
in $true$ or $\mathit{false}$ or, in the worst case, are unchanged. 
For example, 
in Fig.~\ref{ex-litle-slts}, the transition labelled by 
{\small $[\,][G]CompleteTransaction$} and going from 
{\small $Error=\false$} to {\small $Error=\true$} is, 
in Fig.~\ref{ex-raf-slts}, going
from {\small $\StatusWord=ISO\_Ok \land \CurTransaction = None$} to
{\small $\StatusWord\not=ISO\_Ok \land \CurTransaction = None$} with the 
label {\small $[\,][\,]CompleteTransaction$}.
So, its reachability has been made more precise. 
The same effect occurs on transition
{\small $[\,][G]CompleteTransaction$} going from 
{\small $Error=\false$} to {\small $Error=\false$}, 
which is refined by
{\small $[\,][\,]CompleteTransaction$} going from 
{\small $\CurTransaction \not= None$}
to {\small $\CurTransaction = None$} in the super-state 
{\small $Error=\false$}.
These two specializations are directly due to the introduction of 
the {\small $\CurTransaction$} variable.

%
%

\newcommand{\java}{{\bf Java}}
\newcommand{\javacard}{{\bf JavaCard}}
\newcommand{\bom}{{\bf \B OM}}
\newcommand{\bob}{{\bf Bo\B}}
\newcommand{\eden}{{\bf EDEN}}
\newcommand{\cc}{{\bf CC}}
\newcommand{\CC}{{\bf C}ommon {\bf C}riteria}
\newcommand{\toe}{{\bf TOE}}
\newcommand{\jbt}{{\sf jBTools}}        
\newcommand{\cadp}{{\sf CADP}}  
\newcommand{\graphviz}{{\sf GraphViz}}
\newcommand{\uml}{{\bf UML}}
\newcommand{\ocl}{{\bf OCL}}
\newcommand{\jml}{{\bf JML}}

\newcommand{\enable}{\mathit{Enabled}}
\newcommand{\alenable}{\mathit{AlwaysEnabled}}
\newcommand{\crossable}{\mathit{Crossable}}
\newcommand{\alcrossable}{\mathit{AlwaysCrossable}}
\newcommand{\bdef}{~\hat =~}		

\newcommand{\Personnalized}{\mathit{Personalized}}

\section{Verification of Security Properties on \demoney }
\label{verif-security}

   In this section we propose a formalism to express properties relative
   to security aspects and we show how \genesyst\ can be used to verify
   these properties. We will next give a concrete example relative 
   to the \demoney\ case study.

    \subsection{Properties}

   Generally, security is designed and implemented through different
   levels of abstraction. Security policies are defined by a set of rules
   according to which the system can be regulated, in order to guarantee
   expected properties, as confidentiality or integrity. Security
   policies are then implemented through software and hardware
   functions, called security mechanisms. 
   Such an approach has been
   adopted by the \CC\ norm \cite{CC} which proposes,
   through the notion of assurance requirements,
   a catalogue of security policies and a hierarchy of mechanisms.

   In this paper we focus on security properties relative to constraints
   on the global behavior of  the system, as authentication procedures
   or access control. In this case, security requirements can be seen as
   constraints on the execution order of atomic actions, as operation calls. 
   F. Schneider claims in \cite{schneider00enforceable} that
   automata are a well-adapted formalism which can, both, be used to
   specify some forms of security policies and to control implementations during their
   execution.  
   On the other hand, K. Trentelman and M. Huisman \cite{Huisman02} 
   propose a logic that can be used also to express some forms of security 
   properties, as temporal properties on \jml\ specifications.

   We adopt a formalism based on logic formulas, which allows us to point out
   expected behaviors either in specifying correct executions, or in
   specifying security violations. That offers a good
   flexibility and is suitable to describe as well open policies as closed
   policies, respectively relative to negative authorizations and
   positive authorizations \cite{Samarati01AccessControl}. 

      \subsection{Predicates of security properties}
      \label{PredicatesOfSecurityProperties}

   Security properties are often represented as a list of first order logic 
   formulas that have to be verified. We want to define some predicates 
   to make the expression of these formulas easier.
   Predicates that we introduce express the ability of an event to start 
   from a state ($\enable$\ and $\alenable$) and the existence of a 
   transition between two states ($\crossable$ and $\alcrossable$). 

   \begin{defin}[$\enable$, $\alenable$, $\crossable$ and $\alcrossable$]
     \label{DefEnable}
     \label{DefCrossable}
     ~\\
     Given $p_1$ and $p_2$ two state predicates 
     and an event $ev$ from a system $\Sy$ with variables $x$, then:

     \centerline{
       \begin{tabular}{lll}
         $\enable (p_1, ev)$ &          $\bdef$ & $\exists x \cdot (p_1 \logand \guard (ev))$\\
         $\alenable (p_1, ev)$ &        $\bdef$ & $\forall x \cdot (p_1 \Rightarrow \guard (ev))$\\
         $\crossable (p_1, ev, p_2)$ &   $\bdef$ & $\exists x \cdot (p_1 \logand \langle ev \rangle p_2)$\\
         $\alcrossable (p_1, ev, p_2)$ ~~~~~~~& $\bdef$ ~~~~~~~& $\forall x \cdot (p_1 \Rightarrow [ ev ] p_2)$
       \end{tabular}
     }

   \end{defin}

   Let us note that if $\enable (p_1, ev) \logequiv \mathit{false}$, then, 
   for each predicate $p_2$, $\alcrossable (p_1, ev, p_2)$ 
   will be $true$ instead of $\mathit{false}$, which is the intuitive 
value expected. 
   In the same way, if $p_1$ is equivalent to $\mathit{false}$ then $\alenable$ and $\alcrossable$ are always $true$.
   Moreover, we can notice that: 
   
   \centerline{$\crossable (p_1, ev, p_2) \Rightarrow \enable (p_1, ev)$}
   
\noindent
   From this definition we can deduce the properties below, relative to the 
implication:

   \begin{lemme}  
     \label{ProprietesImplication}
     Given $p_1$, $p_2$ and $p_3$ three predicates and an event $ev$ then:
     \begin{itemize}
       \item if $p_1 \Rightarrow p_3$ and $\enable (p_1,ev)$ then $\enable (p_3,ev)$
       \item if $p_3 \Rightarrow p_1$ and $\alenable (p_1,ev)$ then $\alenable (p_3,ev)$
       \item if $p_1 \Rightarrow p_3$ and $\crossable (p_1,ev,p_2)$ then $\crossable (p_3,ev,p_2)$
       \item if $p_2 \Rightarrow p_3$ and $\crossable (p_1,ev,p_2)$ then $\crossable (p_1,ev,p_3)$
       \item if $p_3 \Rightarrow p_1$ and $\alcrossable (p_1,ev,p_2)$ then $\alcrossable (p_3,ev,p_2)$
       \item if $p_2 \Rightarrow p_3$ and $\alcrossable (p_1,ev,p_2)$ then $\alcrossable (p_1,ev,p_3)$
     \end{itemize}
   \end{lemme}

   \noindent
     Here are two examples:

     \noindent {\bf Reactivity of a system.}
     The \javacard\ specification imposes that any APDU instruction is callable 
     at any time. Given $\Sy$ a system and $I$ 
     its invariant, then this formula can be expressed as follows:
     
     \centerline{
       \begin{tabular}{l}
         $\forall ev \cdot (ev \in \interface (\Sy) \Rightarrow \alenable (I, ev))$
       \end{tabular}
     }

\bigskip


     \noindent {\bf Unicity of the ways to reach a state.}
     In some cases, like access control, we want to impose that the only way to reach 
     a state $P$ is to execute a particular event $Begin$. If $I$ 
     is the invariant of $\Sy$, then this property 
     can be expressed as follows:
     
     \centerline{
       \begin{tabular}{l}
         $\forall ev \cdot (ev \in \interface (\Sy) \logand ev\not=Begin \Rightarrow \alcrossable (I, ev, \neg P))$
       \end{tabular}
     }

      \subsection{Property checking using \genesyst\ SLTS}

   Security properties could be verified on \B\ specifications, using 
   definition~\ref{DefEnable}. Nevertheless, in some 
   cases, the SLTS produced by \genesyst\ can be directly exploited.
   Then, the verification consists in using syntactic information relative to 
   enabledness and reachability of transitions.
   Properties~\ref{EnablednessVerificationNonMinimal}--\ref{ReachabilityVerificationMinimal} list the different cases 
   where the predicates above 
   can be directly established from a symbolic labelled transition system.

\newpage
  Properties~\ref{EnablednessVerificationNonMinimal}--\ref{ReachabilityVerificationMinimal} share the following hypothesis:
\noindent 
{\it Given an event $e$ and $q_1$, $q_2$ two states from a SLTS 
     $\slts$, such as $\II(q_1)$ $\not\logequiv$ $\mathit{false}$ and $(q_1,(D,A,e),q_2)$ $\in$ $W_{\slts}$, then predicates $\enable$, $\alenable$, $\crossable$ and 
$\alcrossable$ can be determined as follows:}.

   \begin{lemme}[Enabledness condition - general case] ~\\
     \label{EnablednessVerificationNonMinimal}
   %
     \centerline{
       \begin{tabular}{llll}
  %
         1. & $D \equiv true$                                  & $\Rightarrow$ & $\enable (q_1, e)$\\
         2. & $D \equiv false$                                 & $\Rightarrow$ & $\neg\enable (q_1, e)$\\
         3. & $D \equiv true$ ~~~~~~~~~~~~~~~~~~~~~~~~~~~~~~~~~~~~~~~~~& $\Rightarrow$ & $\alenable (q_1, e)~~~~~~~~ $\\
         4. & $D \equiv false$                                 & $\Rightarrow$~~ & $\neg\alenable (q_1, e)$\\
       \end{tabular}
     }
   \end{lemme}

   If the SLTS used to verify the property is minimal, then 
   Property  \ref{EnablednessVerificationNonMinimal} can be enlarged: 
   the conditions are necessary (and sufficient) and 
   conditions 1 and 4 are refined.

   \begin{lemme}[Enabledness for minimal SLTS]~\\
     \label{EnablednessVerificationMinimal}
     \centerline{
       \begin{tabular}{llll}
  %
         1. & $D \not\equiv false$                             & $\logequiv$   & $\enable (q_1, e)$\\
         2. & $D \equiv false$                                 & $\logequiv$   & $\neg\enable (q_1, e)$\\
         3. & $D \equiv true$ ~~~~~~~~~~~~~~~~~~~~~~~~~~~~~~~~~~~~~~~~~& $\logequiv$   & $\alenable (q_1, e)~~~~~~~~$\\
         4. & $D \not\equiv true$                              & $\logequiv$~~   & $\neg\alenable (q_1, e)$\\
       \end{tabular}
     }
   \end{lemme}

   In the same way, syntactic conditions to check $\crossable$ and $\alcrossable$ predicates are:
   
   \begin{lemme}[Reachability condition - general case] ~
     \label{ReachabilityVerificationNonMinimal}
     
     \centerline{
       \begin{tabular}{llll}
  %
         5. & $A \equiv true$                                  & $\Rightarrow$ & $\crossable (q_1, e, q_2)$\\
         6. & $A \equiv false \logor D \equiv false$           & $\Rightarrow$ & $\neg\crossable (q_1, e, q_2)$\\
         7. & $A \equiv true \logand $                         & &  \\
            & $\forall q_i \cdot (q_2 \not\equiv q_i \Rightarrow (q_1,(D,A_2,e),q_i) \not\in W_{\slts})$ ~~& $\Rightarrow$~~ & $\alcrossable (q_1, e, q_2)$\\
         8. & $A \equiv false $                                & $\Rightarrow$ & $\neg\alcrossable (q_1, e, q_2)$ \\
       \end{tabular}
     }
   \end{lemme}

   Cases 7 and 8 are not symetric, as it would be expected, because, syntacticaly, 
we can just compare names of states, 
   not the intersection of their interpretation.
   Just as for enabledness, the conditions can be enlarged, when the SLTS 
is minimal, as follow:

   \begin{lemme}[Reachability for minimal SLTS] ~
     \label{ReachabilityVerificationMinimal}
     
     \centerline{
       \begin{tabular}{llll}
  %
         5. & $A \not\equiv false$                             & $\logequiv$   & $\crossable (q_1, e, q_2)$\\
         6. & $A \equiv false \logor D \equiv false$   ~~~~~~~~~~~~~~~~~~~~~~& $\logequiv$   ~~& $\neg\crossable (q_1, e, q_2)$\\
  %
  %
         8. & $A \not\equiv true$                              & $\Rightarrow$ & $\neg\alcrossable (q_1, e, q_2)$
       \end{tabular}
     }
   \end{lemme}

   Cases 7 and 8 are just sufficient conditions because of the limitation of the syntactic verification.
   Case 7 is not present in Property~\ref{ReachabilityVerificationMinimal} because it 
is the same as 
   in Property~\ref{ReachabilityVerificationNonMinimal}
%
   Finally, Property~\ref{ProprietesImplication} allows the deduction of derived 
properties from the four properties above, by weakening or strenghtening the states.

    \subsection{Example of a property checking}
    \label{VerificationProprietes}
    
   In this section, we develop a real example of \demoney\ property and we do  
   its verification by using the SLTS given in Figure~\ref{ex-raf-slts}.
   In the \demoney\ specification \cite{DEMONEY-Spec}, 
   the two APDU instructions {\it InitializeTransaction} and {\it CompleteTransaction} 
   have to be executed in sequence, without any other instructions between them
   and without reaching any error state, to make a transaction. 
   However, 
   the card can be withdrawn at any time (modelled by the {\it Reset} event) 
   without generating any error. 
   Transaction atomicity property can be decomposed in five formulas 
   given below, where $I$ stands for the invariant of the \demoney\ specification. 
   Moreover, SLTS of Figure~\ref{ex-raf-slts} is minimal and 
   events are always enabled from all state of the SLTS.
   Finally, note than the invariant $I$ is equivalent to the union of all state 
   predicates (Section \ref{statesC}).
    
   ~
   
   \noindent {\bf Formula 1:} 
       There exists at least a value in $I$ such that the event {\it InitializeTransaction} 
       can reach $\CurTransaction \not= None$:
       
       \centerline{
       \begin{tabular}{l}
         $\crossable (I, InitializeTransaction, \CurTransaction \not= None)$
       \end{tabular}
       }

       \noindent Predicate $\CurTransaction \not= None$ directly corresponds to a 
       state predicate. Since there exists a transition from $\CurTransaction = None \logand \StatusWord=ISO\_Ok$ to 
       $\CurTransaction \not= None$, labelled with $[~][G]InitializeTransaction$, then we 
       can use case~5 of Property~\ref{ReachabilityVerificationMinimal} and conclude that
       the Formula~1 is $true$.

   ~

   \noindent {\bf Formula 2:} 
       For all values, the event {\it InitializeTransaction} goes into 
       the state $\CurTransaction$ $\not=$ $None$ or into an error state:
       
       \centerline{
       \begin{tabular}{l}
         $\alcrossable (I, InitializeTransaction, $\\
         $~~~~\CurTransaction \not= None \logor \StatusWord\not=ISO\_Ok)$
       \end{tabular}
       }

   \noindent $\CurTransaction \not= None$ and $\StatusWord\not=ISO\_Ok$ are two state predicates, 
   and all the transitions labelled with {\it InitializeTransaction} 
   go only in one of these states. Then, due to case~7 of 
   property~\ref{ReachabilityVerificationNonMinimal}, this formula is $true$.

   ~

%
%
%

   \noindent {\bf Formula 3:} 
       From $\CurTransaction\!\!\not=\!\!None$, all events, but $CompleteTransaction$ and $Reset$, go to an error state:
       
       \centerline{
       \begin{tabular}{l}
         $\forall e \cdot (e \in \interface(\Sy) \logand e \not= CompleteTransaction \logand e \not= Reset \Rightarrow $\\
         $~~~~\alcrossable (\CurTransaction \not= None, e, \StatusWord\not=ISO\_Ok)$
       \end{tabular}
       }

   \noindent The two predicates correspond to state predicates and the only events which go elsewhere than 
   $\StatusWord\!\not=\!ISO\_Ok$ from $\CurTransaction\!\not=$ $None$ are {\it CompleteTransaction} 
   and {\it Reset}. Thus Formula 4 is $true$ (case~7 of Property~\ref{ReachabilityVerificationNonMinimal}).

   ~

   \noindent {\bf Formula 4:} 
       Except {\it InitializeTransaction}, no event can reach $\CurTransaction$ $\not= None$:
       
       \centerline{
       \begin{tabular}{l}
         $\forall e \cdot (e \in \interface (\Sy) \logand e\not=InitializeTransaction \Rightarrow$\\
         $~~~~\alcrossable (I, e, \CurTransaction = None))$
       \end{tabular}
       }

   \noindent Predicate $\CurTransaction$ $=$ $None$ is the union of two existing state predicates.
   So, we have to check if there exists an event, different from {\it InitializeTransaction}, that 
   can reach $\CurTransaction \not= None$. Since it is not the case, this formula is $true$
   (case~7 of Property~\ref{ReachabilityVerificationNonMinimal}).

   ~

   \noindent {\bf Formula 5:} 
       No transition labelled by {\it CompleteTransaction} or {\it Reset} is reflexive on 
       state $\CurTransaction$ $\not=$ $None$:
       
       \centerline{
       \begin{tabular}{l}
         $\neg \crossable (\CurTransaction \not= None, CompleteTransaction,$\\
         $~~~~~~~~~~~~~~~\CurTransaction \not= None)$\\
         and $\neg \crossable (\CurTransaction \not= None, Reset, \CurTransaction \not= None)$
       \end{tabular}
       }

   \noindent $\CurTransaction \not= None$ corresponds to a state predicate and no {\it CompleteTransaction} or 
   {\it Reset} reflexive transition occurs. Thus this formula is $true$ (case~5 of 
   Property~\ref{ReachabilityVerificationMinimal}). 

   ~

   The model of \demoney\ is thus correct 
   relatively to the atomicity security property of transactions. 
   However, during the realisation of this example, which is a simplified 
   \demoney\ applet, we found three errors due to an erroneous 
   simplification of our complete model of \demoney. 
   %

   The originality of this approach is to have brought back, under some 
   hypotheses, the verification of security properties to a syntactic 
   checking. 
   However, it is important to be careful about the real
   value of the crossing conditions generated by \genesyst . 
   Indeed, if some proof obligations are not (automatically) discharged, the 
   transitions system will have by-default transitions. Then, to 
   properly exploit the information, we have to be sure that the property 
   to be verify can be checked on a non-minimal SLTS. 


\section{Related works and Conclusion}
\label{conclusion}
\label{related-works}

The work presented here is in line with the ideas presented in
\cite{BC00}, itself inspired by \cite{GS97}. 
In \cite{BC00}, the authors propose the construction of 
a labelled transition system which is a finite state abstraction 
of the behavior of an event-\B\ system. The existence of transitions is 
determined by proof obligations, as here, but the resulting transition 
system does not contain any information about transition crossing.
Moreover, the paper does not consider the refinement step in the 
diagram representation.

Other work is devoted to the translation of dynamic aspects
described by statecharts in the \B\ formalism (for instance 
\cite{Ledang02,SZ02}).
These approaches are inverse of ours, because they go from a 
diagrammatic representation to an encoding in a formal text.
Their objective is to build a \B\ model from UML descriptions.
On our side, we suppose that the model has been stated and we
are interested in representing the precise behavior of the system 
with respect to (a part of) variables, in order to check
properties, or to validate the model against the
requirements.

A similar approach has been envisaged for TLA \cite{TLAInPictures}
and extended in \cite{CMM00,CMM01} to take in account liveness
properties and refinement. As in \cite{BC00}, the generated diagrams
are abstractions of the system behavior.

Several tools are dedicated to the analysis of the behavior of
\B\ components by the way of the animation of machines
\cite{BZTT02} or by local exhaustive model checking \cite{LB03}. 
Even if some of them allow the generation of symbolic
traces, these tools can be considered as ``testing'' tools. They provide
particular execution sequences of the system, not a static representation
of all the behaviors. In \cite{VT03}, the authors describe the generation of 
statecharts from event-\B\ systems, but their approach suffers from several
restrictions and their diagrams are not symbolic.

In this paper, we have presented the \genesyst\ tool, its logical 
foundations and its application to the verification of security properties.
In the first part, we introduced the definition of traces of event-\B\ systems
and refinements. We formalized the notion of symbolic labelled transition 
systems, with transitions decorated by enabledness and reachability 
predicates. This gives a complete and precise view of the behavior 
of the system, which can be exploited for various objectives.

We described the algorithm that is implemented to generate a
SLTS from a \B\ system and a set of states, characterized by predicates.
The computation of effective transitions between states is performed by 
proving proof obligations. Due to the indecidability of the proof process,
we have the choice between two kinds of (non exclusive) results: the
generation is automatic, but we can get more transitions than in the
real system, or the user completes interactively the non-conclusive proofs
and then, the resulting automaton reflects exactly the behavior of 
the system.

The user can take profit of the freedom degree achieved by the choice
of the states, to obtain the best analyses useful for him/her purpose.
Non classical verification techniques can be designed and implemented
at this stage, to assess or to validate the model, as it was shown 
in the last part of the paper. This opens a large field of research
in the domains of security properties, confidentiality, access control,
validation of models with respect to the requirements, automatic 
documentation of specifications, etc.
Our present research work is to develop a set of techniques in the 
GECCOO\footnote{``G\'en\'eration de Code Certifi\'e Orient\'e Objet''.
Project of Program ``ACI S\'ecurit\'e Informatique'', 2003.}
project to express and to check security properties,
as it was sketched in the paper. We want to investigate
the extraction of states from the specification of property automata, 
the use of refinement to split states and achieve a suitable
level of decomposition in order to check a property. Another work is to 
deal with complex \B\ models (several refinement chains together with 
composition clauses \sees, \incl, etc.), either by composing partial labelled
transition systems, or by flattening a structured model before
computing the whole associated SLTS.

\bibliographystyle{plain}
\bibliography{demoney,RelatedWorks,Bsystem}

\begin{thebibliography}{10}

\bibitem{BBook}
J.-R. Abrial.
\newblock {\em The {B} Book - Assigning Programs to Meanings}.
\newblock Cambridge University Press, August 1996.

\bibitem{abrial96}
J.-R. Abrial.
\newblock {Extending {{B}} without Changing it (for Developing Distributed
  Systems)}.
\newblock In H.~Habrias, editor, {\em First {{B}} conference}, Putting into
  Practice Methods and Tools for Information System Design, IRIN, pages
  169--191, 1996.

\bibitem{JRAbrialB98}
J.R. Abrial and L.~Mussat.
\newblock {Introducing Dynamic Constraints in {B}}.
\newblock In D.~Bert, editor, {\em B'98: Recent Advances in the Development and
  Use of the B Method}, LNCS 1393, pages 83--128. Springer-Verlag, 1998.

\bibitem{BZTT02}
F.~Ambert, F.~Bouquet, S.~Chemin, S.~Guenaud, B.~Legeard, F.~Peureux,
  M.~Utting, and N.~Vacelet.
\newblock {BZ-testing tools: A tool-set for test generation from Z and B using
  constraint logic programming}.
\newblock In {\em Formal Approaches to Testing of Software (FATES'02)}, pages
  105--120. INRIA, 2002.

\bibitem{BC00}
D.~Bert and F.~Cave.
\newblock {Construction of Finite Labelled Transition Systems from B Abstract
  Systems}.
\newblock In W.~Grieskamp, T.~Santen, and B.~Stoddart, editors, {\em Integrated
  Formal Methods}, LNCS 1945, pages 235--254. Springer-Verlag, 2000.

\bibitem{CMM00}
D.~Cansell, D.~M\'ery, and S.~Merz.
\newblock {Predicate Diagrams for the Verification of Reactive Systems}.
\newblock In W.~Grieskamp, T.~Santen, and B.~Stoddart, editors, {\em Integrated
  Formal Methods}, LNCS 1945, pages 380--397. Springer-Verlag, 2000.

\bibitem{CMM01}
D.~Cansell, D.~M\'ery, and S.~Merz.
\newblock {Diagram Refinements for the Design of Reactive Systems}.
\newblock {\em Journal of Universal Computer Science}, 7(2), 2001.

\bibitem{CC}
Common Criteria.
\newblock {\em Common Criteria for Information Technology Security Evaluation,
  Norme ISO 15408 - version 2.1}, Aout 1999.

\bibitem{GS97}
S.~Graf and H.~Sa\"{\i}di.
\newblock {Construction of Abstract State Graphs with PVS}.
\newblock In {\em Computer-Aided Verification (CAV'97)}, LNCS 1254.
  Springer-Verlag, 1997.

\bibitem{csp}
C.A.R. Hoare.
\newblock {\em {Communicating Sequential Processes}}.
\newblock Prentice Hall, 1985.

\bibitem{lamport94}
L.~Lamport.
\newblock {A} {T}emporal {L}ogic of {A}ctions.
\newblock {\em ACM Transactions on Programming Languages and Systems},
  16(3):872--923, may 1994.

\bibitem{TLAInPictures}
L.~Lamport.
\newblock {TLA} in {P}ictures.
\newblock {\em Software Engineering}, 21(9):768--775, 1995.

\bibitem{Ledang02}
H.~Ledang and J.~Souqui\`eres.
\newblock {Contributions for Modelling UML State-charts in B}.
\newblock In M.~Butler, L.~Petre, and K.~Sere, editors, {\em IFM}, LNCS 2335,
  pages 109--127. Springer-Verlag, 2002.

\bibitem{LB03}
M.~Leuschel and M.~Butler.
\newblock {ProB: A Model Checker for B}.
\newblock In K.~Akari, S.~Gnesi, and D~Mandrioli, editors, {\em FME 2003:
  Formal Methods}, LNCS 2805, pages 855--874. Springer-Verlag, 1997.

\bibitem{DEMONEY-AnnexeDetaillees}
R.~Marlet.
\newblock {DEMONEY: Java Card Implementation}.
\newblock Public technical report, SECSAFE project, 11 2002.

\bibitem{DEMONEY-Spec}
R.~Marlet and C.~Mesnil.
\newblock {DEMONEY : A demonstrative Electronic Purse - Card Specification -}.
\newblock Public technical report, SECSAFE project, 11 2002.

\bibitem{Samarati01AccessControl}
P.~Samarati and S.~De~Capitani di~Vimercati.
\newblock {Access Control: Policies, Models, and Mechanisms}.
\newblock In {\em Revised versions of lectures given during the IFIP WG 1.7
  International School on Foundations of Security Analysis and Design on
  Foundations of Security Analysis and Design}, pages 137--196.
  Springer-Verlag, 2001.

\bibitem{schneider00enforceable}
F.~B. Schneider.
\newblock Enforceable security policies.
\newblock {\em Information and System Security}, 3(1):30--50, 2000.

\bibitem{SecSafe}
SecSafe.
\newblock {SecSafe Porject Home Page}.
\newblock http://www.doc.ic.ac.uk/~siveroni/secsafe/.

\bibitem{SZ02}
E.~Sekerinski and R.~Zurob.
\newblock {Translating Statecharts to B}.
\newblock In M.~Butler, L.~Petre, and K.~Sere, editors, {\em IFM}, LNCS 2335,
  pages 128--144. Springer-Verlag, 2002.

\bibitem{APIJavaCard}
SUN.
\newblock {Java Card 2.1 Platform Specifications}.
\newblock http://java.sun.com/products/javacard/specs.html.

\bibitem{Huisman02}
K.~Trentelman and M.~Huisman.
\newblock {Extending JML Specifications with Temporal Logic}.
\newblock In {\em {Algebraic Methodology And Software Technology (AMAST '02)}},
  LNCS 2422, pages 334--348. Springer-Verlag, 2002.

\bibitem{VT03}
J.-C. Voisinet and B.~Tatibouet.
\newblock {Generating Statecharts from B Specifications}.
\newblock In {\em 16th Int Conf. on Software and System Engineering and their
  applications (ISCEA 2003)}, volume~1, 2003.

\end{thebibliography}
                        
\newpage
\section*{Appendix}
Machine of the \demoney\ specification (diagram in Fig.~\ref{ex-litle-slts}, Section~\ref{ts-demoney-1}):


\newcommand{\bimpl}{\logimpli}

\begin{small}
\machinesbox{
\>\mach\ $Demoney$\\
\>\var\ \\
\>\>$\Error, \TransEngage$\\
\>\inv\ \\
\>\>$\Error \in \bool \logand \TransEngage \in \bool \logand $ \\
\>\>$(\Error=\true \bimpl \TransEngage=\false) \logand$\\
\>\>$(\TransEngage=\true \bimpl \Error=\false)$\\
\>\assert\ \\
\>\>/* The assertion provides the states for tool \genesyst\ */\\
\>\>/* Here, only two states are considered according to the $\Error$ values */~~\\
\>\>$\Error=\false \logor \Error=\true$\\
\>\init\ \\
\>\>$\Error := \false ~||~ \TransEngage := \false$\\
\>\oper\\
\>\>$\Reset = $ \bbegin\ $\TransEngage:=\false ~||~ \Error:=\false$ \bend;~~~~\\
\>\>$\GetData =$\\
\>\>\>\bif\ $\TransEngage=\true$ \bthen\ \\
\>\>\>\>$\Error:=\true ~||~ \TransEngage:=\false$\\
\>\>\>\belse\ $\Error:=\false$\\
\>\>\>\bend;\\
\>\>$\InitializeTransaction =$\\
\>\>\>\bif\ $\TransEngage=\true$ \bthen\\
\>\>\>\>$\Error:=\true ~||~ \TransEngage:=\false$\\
\>\>\>\belse\\
\>\>\>\>\bany\ $SW$ \bwhere\ $SW \in \bool$ \bthen\\
\>\>\>\>\>$\Error:=SW ~||~ \TransEngage := \bbool (SW=\false)$\\
\>\>\>\>\bend \\
\>\>\>\bend ;\\
\>\>$\CompleteTransaction =$ \\
\>\>\>\bif\ $\TransEngage=\false$ \bthen\ \\
\>\>\>\>$\Error:=\true$\\
\>\>\>\belse\ $\Error:=\false ~||~ \TransEngage := \false$\\
\>\>\>\bend \\
\>\bend\\
}
\end{small}

\bigskip
\noindent
Refinement of the \demoney\ specification (diagram in Fig.~\ref{ex-raf-slts}, Section~\ref{ts-demoney-2}):


\begin{small}
\machinesbox{
\>\refi\ $Demoney\_R1$\\
\>\refines\ $Demoney$\\
\>\sets\ \\
\>\>$\TypeTransactions = \{Credit,Debit,None\} ;$\\
\>\>$\TypesOfStatus = \{ISO\_Error, ISO\_Ok\}$\\
\>\var\ \\
\>\>$\StatusWord, \CurTransaction, \ChannelIsSecured$~~~~~~~~~~~~~~~~~~~~~~~~~~~~~~\\
}

\machinesbox{
\>\inv\\
\>\>$\StatusWord \in \TypesOfStatus \logand \CurTransaction \in \TypeTransactions \logand$ ~~~~\\
\>\>$\ChannelIsSecured \in \bool \logand$ \\
\>\>$((\StatusWord=ISO\_Ok) \logequiv (\Error=\false)) \logand$ \\
\>\>$((\TransEngage=\true) \logequiv (\CurTransaction \not= None)) \logand$ \\
\>\>$((\CurTransaction \not= None) \bimpl \ChannelIsSecured=\true) \logand$ \\
\>\>$((\StatusWord \not=ISO\_Ok) \bimpl (\CurTransaction = None))$ \\
\>\assert\\
\>\>/* Each abstract state is decomposed in two concrete states */\\
\>\>/* One of these states is not reachable */\\
\>\>$((\Error=\true) \logequiv$ \\
\>\>\>$((\StatusWord\not=ISO\_Ok \logand \CurTransaction=None)$ \\
\>\>\>$\logor (\StatusWord\not=ISO\_Ok \logand \CurTransaction\not=None)))$\\
\>\> $\logand $\\
\>\>$((\Error=\false) \logequiv$ \\ 
\>\>\>$((\StatusWord=ISO\_Ok \logand \CurTransaction=None)$ \\ 
\>\>\>$\logor (\StatusWord=ISO\_Ok \logand \CurTransaction\not=None)))$\\
\>\init\\
\>\>$\StatusWord := ISO\_Ok ~||~\ChannelIsSecured := \false ~||$\\
\>\>$\CurTransaction := None$ \\
\>\oper\\
\>\>$\Reset$ = \bbegin \\
\>\>\>\>$\StatusWord:=ISO\_Ok ~||~\ChannelIsSecured := \false ~||$\\
\>\>\>\>$\CurTransaction := None$\\
\>\>\>\bend ;\\
\>\>$\GetData$ = \\
\>\>\>\bif\ $\CurTransaction\not=None$ \bthen\\
\>\>\>\>$\StatusWord:=ISO\_Error ~||~ \CurTransaction:=None$\\
\>\>\>\belse\\
\>\>\>\>$\StatusWord:=ISO\_Ok$\\
\>\>\>\bend ;\\
\>\>$\InitializeTransaction$ = \\
\>\>\>\bif\ $\CurTransaction\not=None \logor \ChannelIsSecured = \false$ \bthen\\
\>\>\>\>$\StatusWord:=ISO\_Error ~||~ \CurTransaction:=None$\\
\>\>\>\belse\\
\>\>\>\>$\StatusWord :\in \TypesOfStatus ;$\\
\>\>\>\>\bif\ $\StatusWord=ISO\_Ok$ \bthen\\
\>\>\>\>\>$\CurTransaction :\in \{Debit,Credit\}$\\
\>\>\>\>\bend\\
\>\>\>\bend ;\\
\>\>$\CompleteTransaction$ = \\
\>\>\>\bif\ $\CurTransaction=None$ \bthen\ \\
\>\>\>\>$\StatusWord:=ISO\_Error$\\
\>\>\>\belse\ \\
\>\>\>\>$\CurTransaction := None ~||~ \StatusWord:=ISO\_Ok$ \\
\>\>\>\bend\\
\>\bend\\
}
\end{small}

\end{document}